# What Can Heterogeneity Add to the Scientometric Map?
# Steps towards algorithmic historiography [1]

Loet Leydesdorff

Amsterdam School of Communications Research (ASCoR), University of Amsterdam

Kloveniersburgwal 48, 1012 CX Amsterdam, The Netherlands;

loet@leydesdorff.net ; http://www.leydesdorff.net

**Introduction**

The Actor Network represents heterogeneous entities as *actants* (Callon *et al*., 1983; 1986). Although computer programs for the visualization of social networks increasingly allow us to represent heterogeneity in a network using different shapes and colors for the visualization, hitherto this possibility has scarcely been exploited (Mogoutov *et al*., 2008). In this contribution to the *Festschrift*, I study the question of what heterogeneity can add specifically to the visualization of a network. How does an integrated network improve on the one-dimensional ones (such as co-word and co-author maps)? The *œuvre* of Michel Callon is used as the case materials, that is, his 65 papers which can be retrieved from the (*Social*) *Science Citation Index* since 1975.[2]

**Methods**

My methods are standard and straightforward. Author names, the names of the respective journals, the titles, the references, etc., can all be attributed to documents as units of analysis. I construct a matrix with the 65 retrieved documents as the cases, 48 unique co-authors of Callon as variables, and the 27 words which occurred more than twice in the titles of these documents as another set of variables.

---

[1] in: *Festschrift for Michel Callon's 65th birthday*, Madeleine Akrich, Yannick Barthe, Fabian Muniesa, and Philip Mustar (Eds.). Paris: École Nationale Supérieure des Mines (forthcoming).

[2] Using Google Scholar, 992 papers can be retrieved using the name of "M Callon" as a search string. The latter set contains 501 unique co-authorship relations. If one would add to this the numerous journals and words contained in this set, the visual representation would rapidly become unreadable. For this reason, I chose to use the 65 papers in the (*Social*) *Science Citation Indices*.

The papers appeared in 26 journals during the period 1975 – 2009.[3] These journal names are added as a third set of variables. The number of variables therefore is (48 + 27 + 26 =) 101. The matrix is normalized in terms of co-occurrences among the variables using the cosine for the similarity (Ahlgren *et al.*, 2003; Leydesdorff, 2008). The figures are drawn using the spring-based algorithm of Kamada and Kawai (1989) as available in Pajek for the visualization.[4] The size of the nodes is in proportion to the logarithm of the frequency of occurrence in the data.

**Results**

Figure 1 provides a full representation including the three relevant sets of variables.

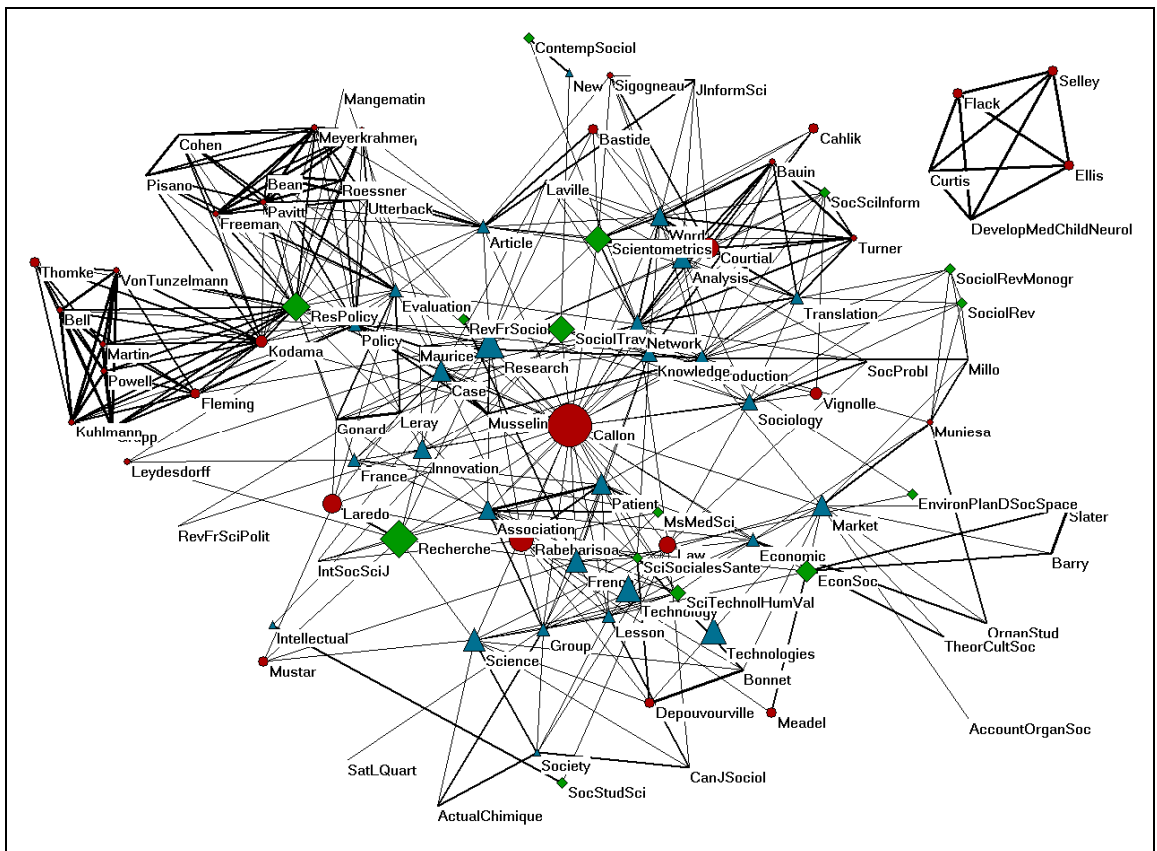

**Figure 1**: Integrated map of 48 authors, 27 words, and 26 journals based on 65 publications of Michel Callon contained in the ISI database; cosine ≥ 0.2 (words: ▲; authors: ●; journals: ♦).

---

[3] The records were downloaded on June 9, 2009.
[4] Pajek is a freeware program for network visualization available at http://vlado.fmf.uni-lj.si/pub/networks/pajek/ .

Figure 1 shows that Callon entertains (or entertained) several strong network components. Two of them (at the top left side) are related to the journal *Research Policy*. One group is based on co-authored editorials, and the other on two evaluative studies of the development of this journal in 1993 and 1999. The isolated group on the top-right side is based on co-authorship relations in a single paper entitled "Ultrasonographic Study of Sucking and Swallowing by Newborn-Infants" (*Developmental Medicine and Child Neurology* 28(6) (1986) 821-823.) These three clusters are not so rich in terms of the prevailing semantics in the title words of Callon's publications in other parts of the map.

The words (triangles) are more densely connected in the center of the figure and the two remaining groups. One cluster is with Courtial as a main co-author on the top side, publishing mainly in *Scientometrics* and *Social Science Information* about co-word and network analysis. The other group is oriented more locally with Volonola Rahebarisoa and John Law as Callon's main co-authors. This is the research group focusing on patient associations in France. Had we used a French database or the *Google Scholar* data, this cluster would have been much more pronounced. In addition to French journals, however, publications appear in international journals such as *Science, Technology & Human Values, Economy and Society,* etc.

Let us first compare this extremely rich representation with the co-author map of these same papers in Figure 2 and the co-word map in Figure 3.

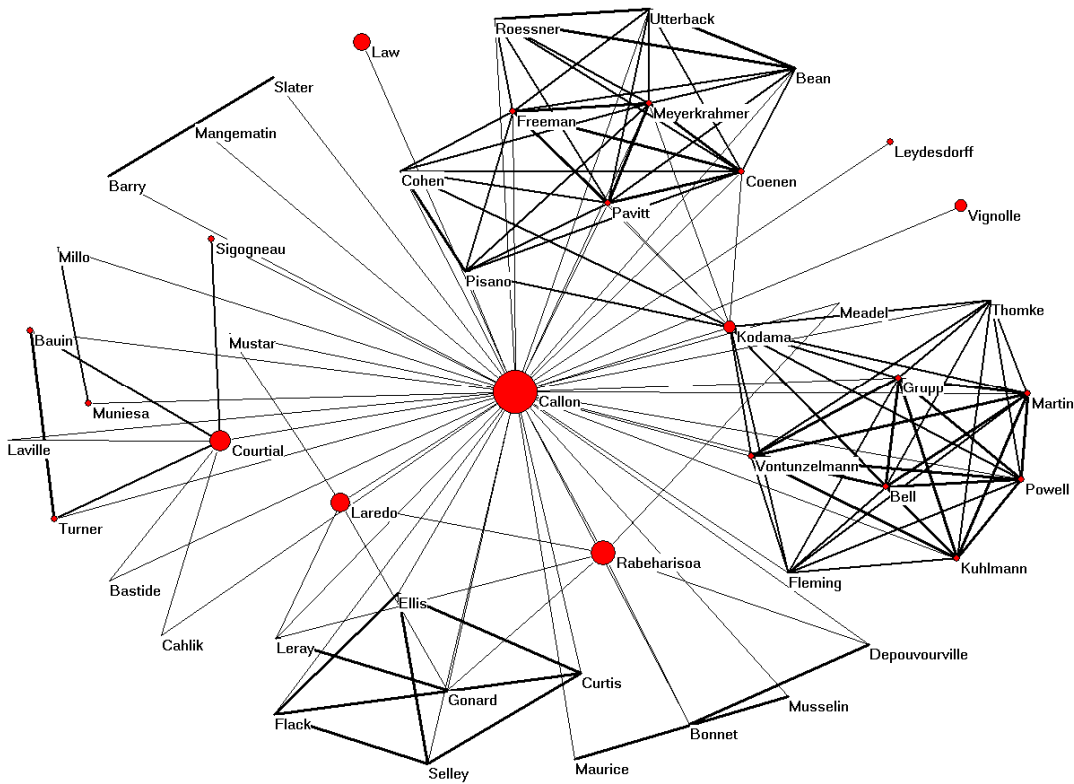

**Figure 2**: Map of 48 co-authors of Michel Callon; $N = 65$; no threshold.

The clusters with relatively less semantic content in Figure 1 dominate Figure 2. The node representing Fumio Kodama provides an articulation point between the two groups publishing in *Research Policy*. The groups with the rich semantics are less pronounced, although one can retrieve them if one knows the subject of the respective collaborations. Philippe Larédo is more important in connecting the two groups from a policy perspective than he was in the previous representation. However, the research lines developed with Jean-Pierre Courtial and Volonola Rabeharisoa are not connected by co-authorship relations other than via Callon himself. In summary, this representation of co-authorship relations informs us mainly about social relations. The cognitive dimensions of these collaborations remain latent.

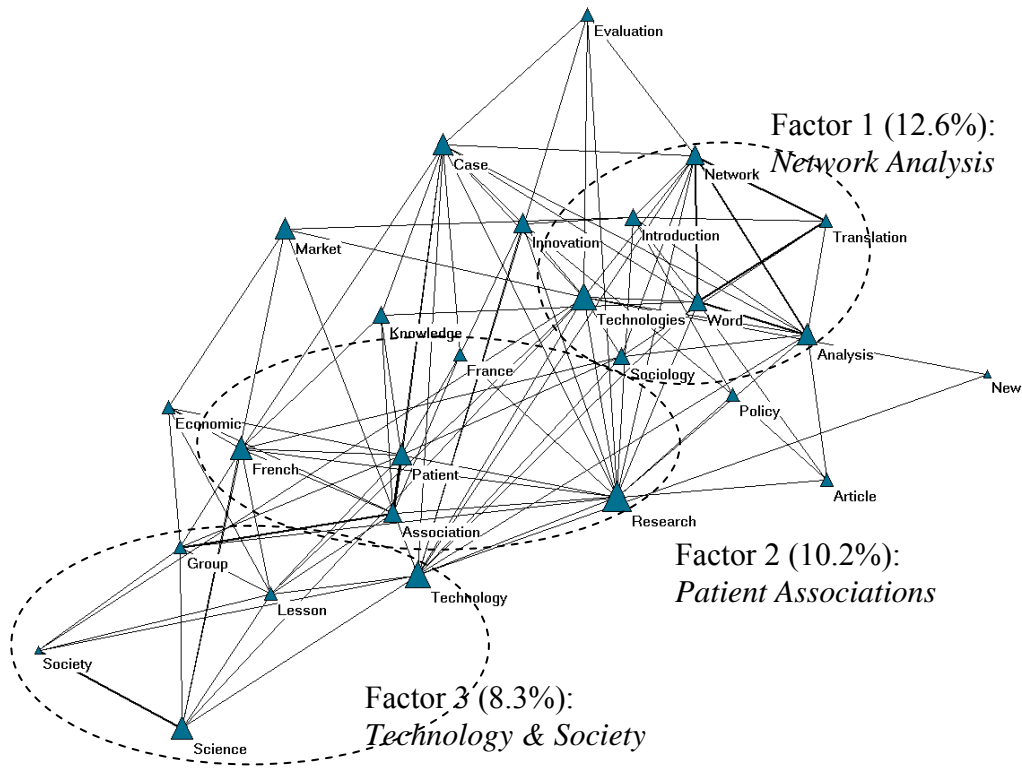

**Figure 3**: Co-word map of title words in 65 papers of Michel Callon; no threshold.

Figure 3 is based on the 27 words which occurred more than twice in the 65 publications under study. Words and their co-occurrences (co-words) are less codified than citation relations (Leydesdorff, 1989). The structure in the data is therefore less pronounced than with citation or co-authorship relations. One may need additional (statistical) analysis to distinguish the groupings clearly. In Figure 3, the three main factors are circled for the sake of clarification. Words in these three components correspond to three of Callon's main research interests. However, the three factors explain only 31.1% of the variance contained in the datamatrix.

Both the co-author and co-word maps thus are relatively uninformed when compared with the integrated map in Figure 1, with the journals also added. One needs additional information—for example, from factor analysis—to understand the structure of the semantic map. The co-author map is easier to understand in terms of institutional affiliations, but this perspective is not informative without local knowledge about the cognitive agendas which motivated these authors to collaborate.

**The Evolution of Callon's *Oeuvre***

The static representations cannot teach us anything about the evolution of the research trajectory of the author. Figure 4 provides the breakdown of Figure 1 in six periods of five years, that is, 1975-2005. Recently, these figures can also be animated using, for example, the dynamic version of Visone (at [http://www.leydesdorff.net/callon/animation](http://www.leydesdorff.net/callon/animation);[5] cf. Leydesdorff & Schank, 2008).

Figure 5 (which for technical reasons cannot be made fit into Figure 4) provides the last period 2005-2009 as an example. Note that although important to Callon's role in organizing the field of science and technology studies, the relation with *Research Policy* did not lead to new words entering his repertoire. The new words (in green) are introduced on the right side of the picture in relation to sociological journals. This is also the case for other years: the relation with the editorial board of *Research Policy* does not play a role in the period 2000-2005 in terms of publications or coauthorship relations.

---

[5] Different from Figures 1 to 3, the animations are based on title words, author names, and journal names that occur more than once; cosine $\geq 0.2$.

**Figure 4**: Evolution of the integrated map of figure 1 in five-year time steps (1975-2005); title words, author names, and journal names which occur more than once;[4] cosine ≥ 0.2.

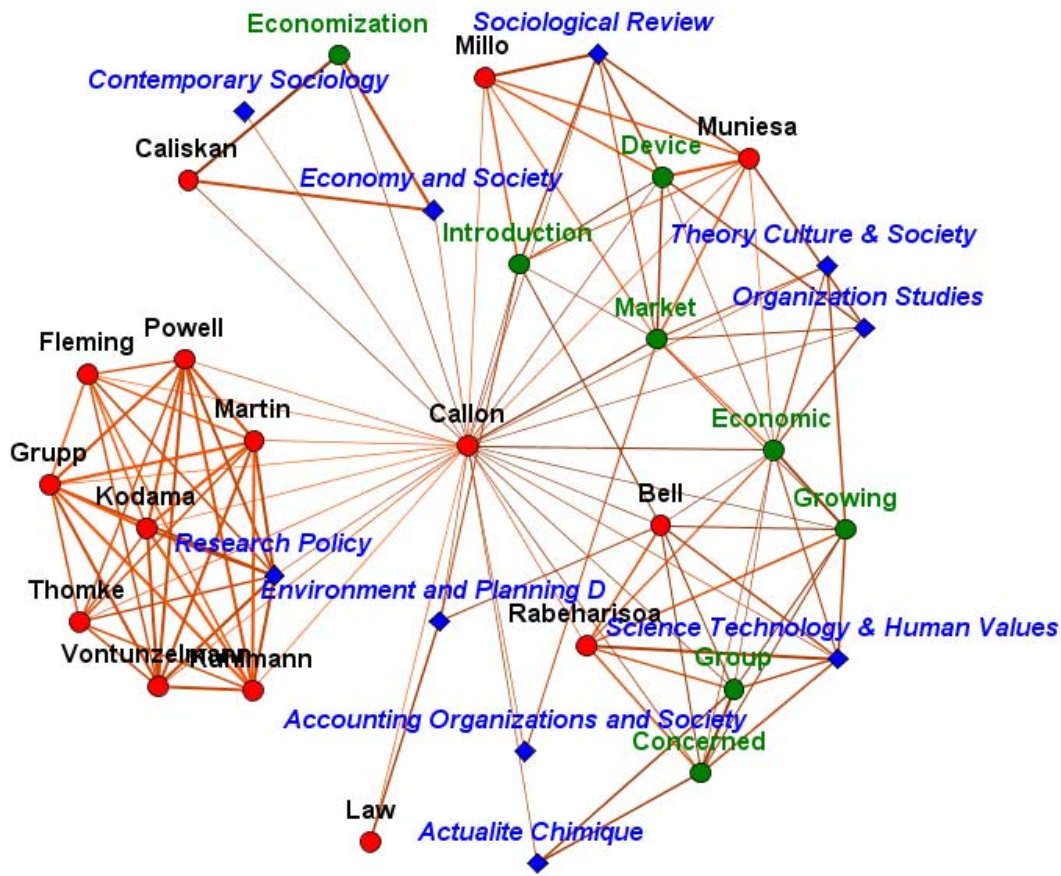

**Figure 5**: Callon's repertoire, coauthorship relations, and publication outlets during the period 2005-2009; title words, author names, and journal names which occur more than once; cosine ≥ 0.2.

*Scientometrics* which was a major focus of attention in the earlier period (1985-1990), completely disappeared from the screen after 1995. Vololona Rabeharisoa—whom I first met as a PhD student in 1990—introduced the focus on medical technologies and patient organizations after 1995. The general issue of how technology transforms society and its economy becomes gradually more pronounced during Callon's career, but the contributions are less often co-authored.

**Conclusion**

In summary, Michel Callon was right when he hypothesized that one has to combine the information contained in the various maps in order to obtain a meaningful and rich representation. Author names contribute to the *semiosis* in actor networks. Social and cognitive structures are interwoven into textual domains. Unlike social network analysis, with its main focus on agents, scientometrics is interested not only in the social structures but also in understanding the semantic map (Callon *et al*., 1993). Conversely, the cognitive constructs (e.g., clusters of words) can inform the appreciation of social relations. Adding the journals further enriches this map as any other relevant category might do (e.g., institutional affiliations). Further interpretation may increasingly lead to the development of algorithmic historiography (Garfield *et al*., 2003) as a field which Callon and his colleagues (1983 and 1986) have envisaged.